\documentclass[aps,prl,reprint,showpacs,twocolumn,superscriptaddress,floatfix]{revtex4-1}
\usepackage{graphicx}
\usepackage{amssymb}
\usepackage{amsmath}
\usepackage{extarrows}

\begin{document}

\title{Over-Bias Light Emission due to Higher Order Quantum Noise of a Tunnel Junction}

\author{F. Xu}

\affiliation{Fachbereich Physik, Universit\"at Konstanz, D-78457 Konstanz, Germany}

\author{C. Holmqvist}

\affiliation{Fachbereich Physik, Universit\"at Konstanz, D-78457 Konstanz, Germany}

\author{W. Belzig}

\affiliation{Fachbereich Physik, Universit\"at Konstanz, D-78457 Konstanz, Germany}

\email[]{Wolfgang.Belzig@uni.kn}

\date{\today}

\begin{abstract}
Understanding tunneling from an atomically sharp tip to a metallic surface requires to account for  interactions on a nanoscopic scale. Inelastic tunneling of electrons generates emission of photons, whose energies intuitively should be limited by the applied bias voltage. However, experiments [Phys. Rev. Lett. \textbf{102}, 057401 (2009)] indicate that more complex processes involving the interaction of electrons with plasmon polaritons lead to photon emission characterized by over-bias energies. We propose a model of this observation in analogy to the dynamical Coulomb blockade, originally developed for treating the electronic environment in mesoscopic circuits. We explain the experimental finding quantitatively by the correlated tunneling of two electrons interacting with an LRC circuit modeling the local plasmon-polariton mode. To explain the over-bias emission, the non-Gaussian statistics of the tunneling dynamics of the electrons is essential.

\end{abstract}



\maketitle

Light emission of electrons tunneling from a scanning tunneling microscope (STM) to a metallic surface has already been studied for many years \cite{Berndt:90}. The basic mechanism leading to light emission has been identified as the interaction of the tunneling electrons with a localized surface plasmon-polariton (SPP) mode \cite{Berndt:91}. Considering such a mechanism in a simple picture shows that the emitted light spectrum is limited in frequency by the bias voltage according to $\hbar\omega<eV$. This is a simple consequence of the presence of Fermi seas in the electronic leads, which prohibits inelastic tunneling transitions with higher energy exchange due to the Pauli principle. The SPP resonance, which is finally responsible for the photon emission, acts as a frequency filter and hence the measured spectrum is essentially the SPP resonance cut off at a frequency $eV/\hbar$. This picture has been confirmed in numerous
experimental \cite{Lambe:76} and theoretical \cite{Rendell:78} studies.
However, a closer look at some experiments \cite{Hoffmann:03,Schull:09,Schneider:10} reveals the unexpected feature that, in addition, light with energy $\hbar\omega>eV$ is emitted that shows a spectrum which is also reminiscent of the SPP modes.
 Using energy considerations, such a process can be attributed to two simultaneously tunneling electrons providing enough energy to explain the observation of over-bias emission.
However, why the electrons tunnel in a correlated manner remains speculative. A possible explanation is a hot electron distribution created by an effective electron-electron interaction \cite{Gadzuk:71,Schneider:13}, which however has been not experimentally tested yet.

In this Letter, we will develop an alternative approach based on the idea that on a short time scale multi-electron coherent processes appear at a tunnel junction. In a coherent two-electron tunneling process where each electron gives contributes an energy $\lesssim eV$ an excitation of an overbias plasmon resonance via a virtual state can be created. The SPP mode finally leads to the over-bias light in a standard emission process. Essentially in our model the coupled electron-SPP system has to be treated as a quantum coherent entity since intermediate virtual states are involved. Considering a single sufficiently damped SPP resonance, we can quantitatively reproduce the experimentally observed bias voltage-dependent emission spectrum.

\begin{figure}
\includegraphics[width=0.99\columnwidth,clip=true]{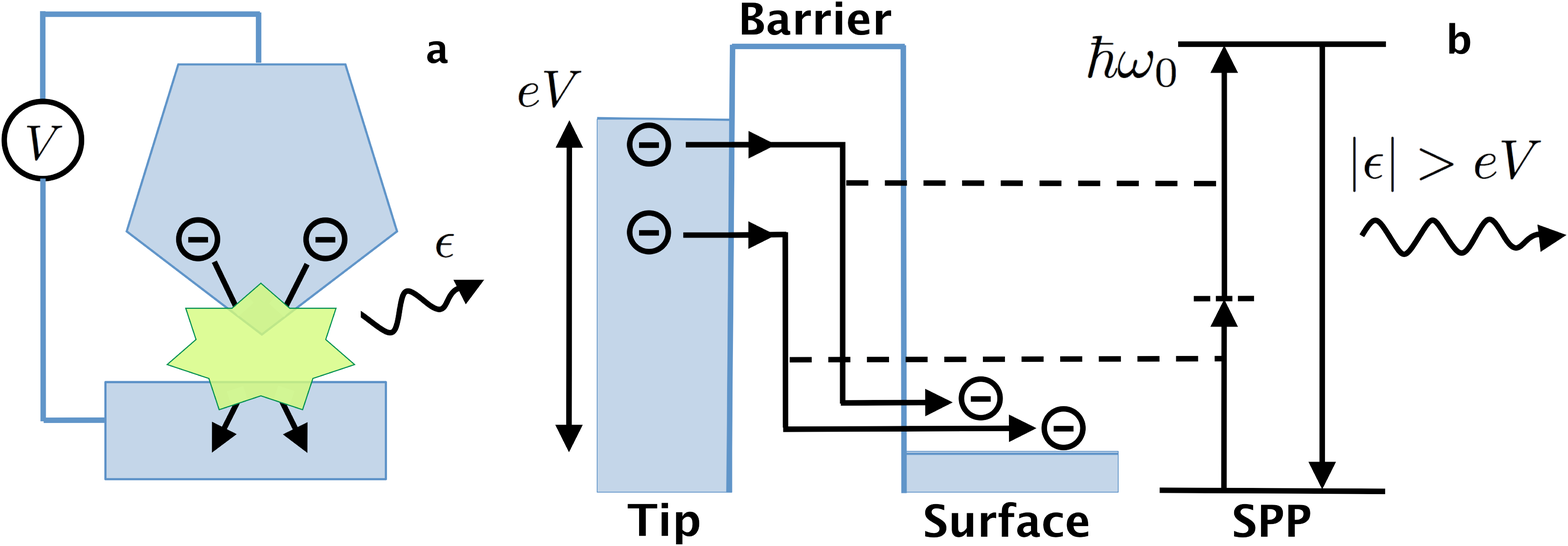}\\
\includegraphics[width=0.45\columnwidth,clip=true]{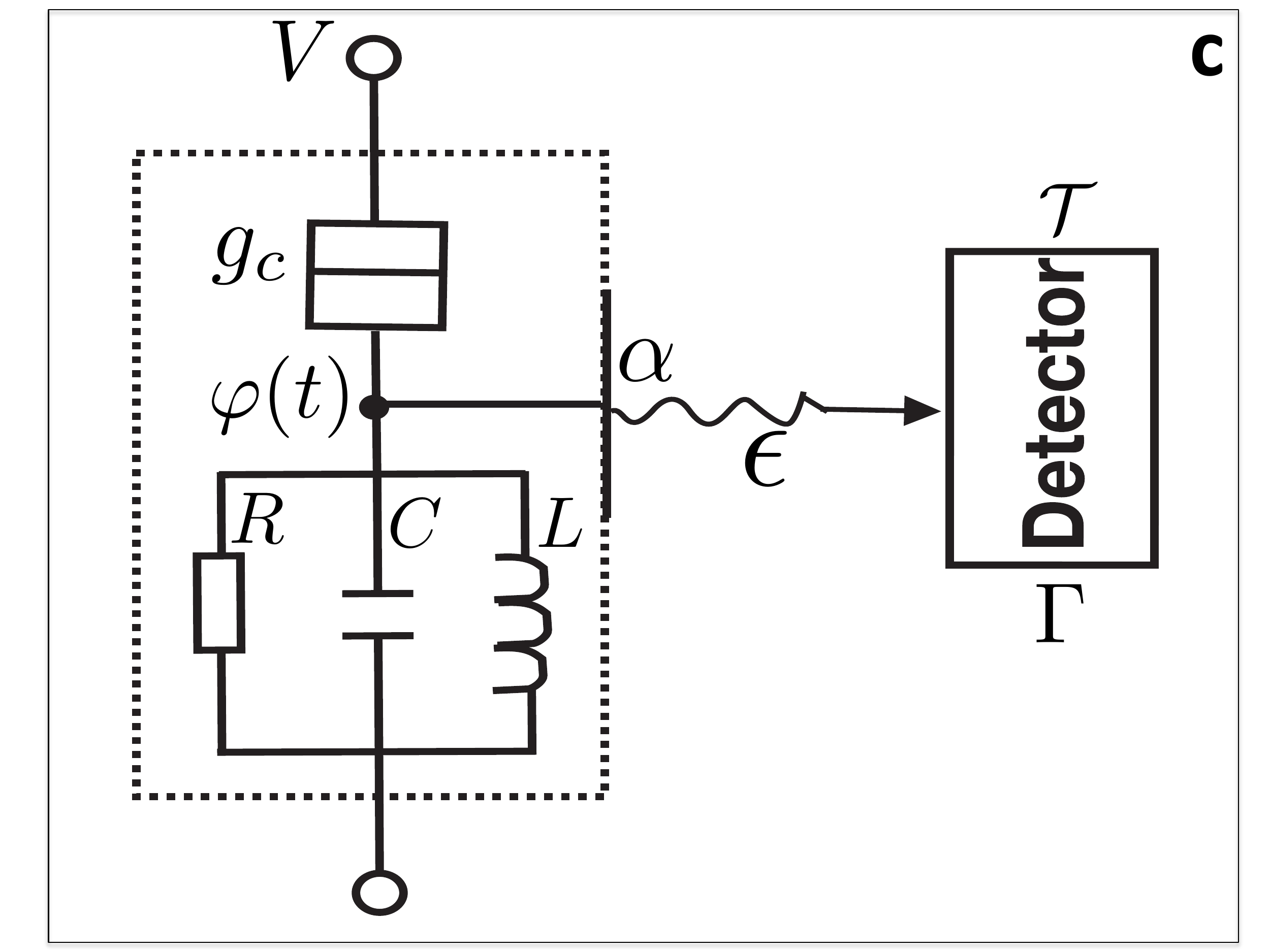}
\includegraphics[width=0.45\columnwidth,clip=true]{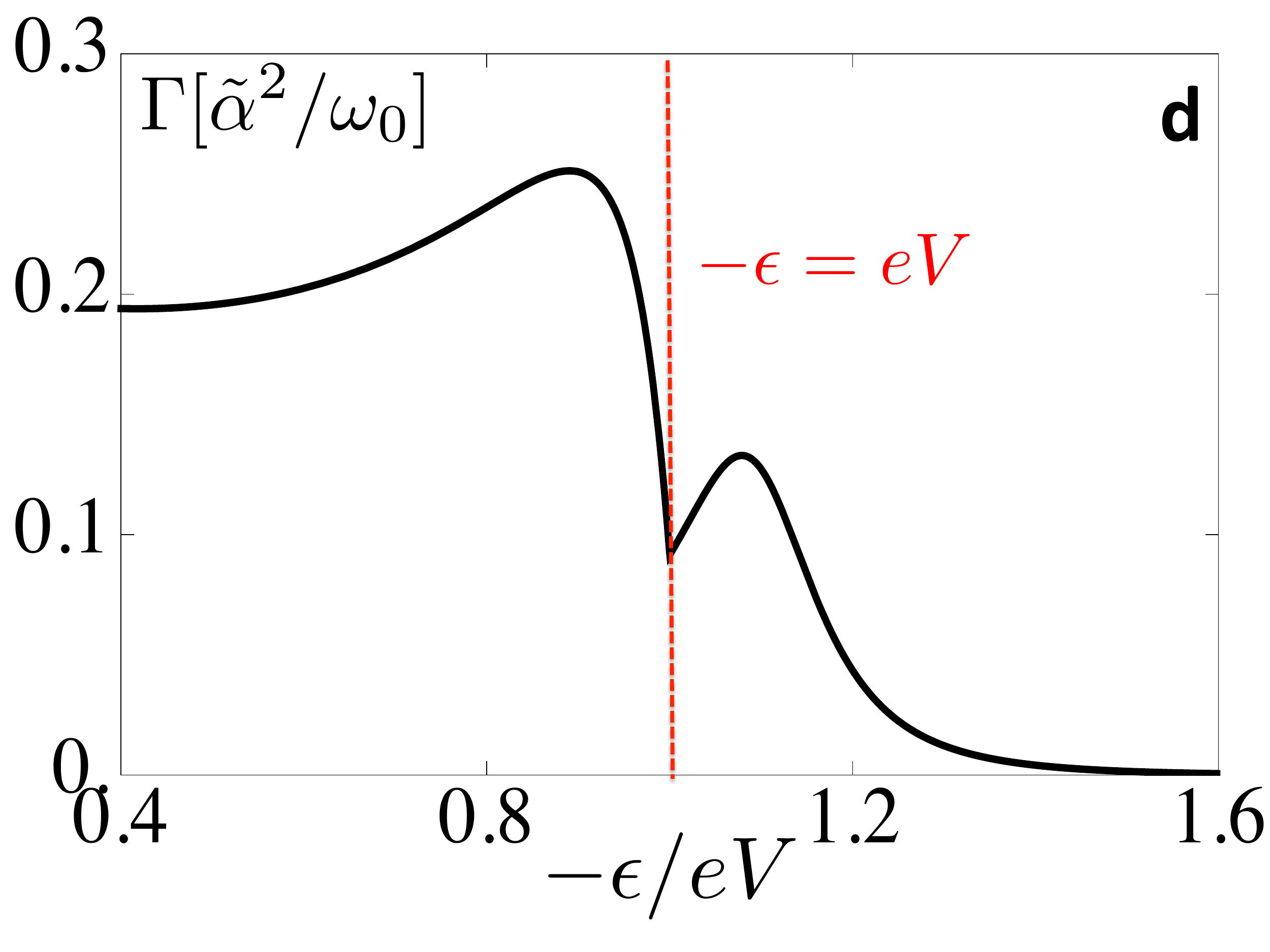} 
\caption{a) Sketch of an STM contact with bias voltage $V$ showing the correlated two-electron tunneling process. 
The electrons interact via an SPP mode (green), 
that enhances the light emission from the STM junction \cite{Ebbesen:03}.
b) The electron tunneling process in energy space shows how two electrons excite an SPP via a virtual state. The SPP decays by emitting a photon with an energy $|\epsilon|>eV$.  c) The electromagnetic model circuit: an LRC resonant circuit mimics the (damped) SPP and the photons emitted from the coupled tunnel junction are captured by the detector. d) The theoretical detection rate $\Gamma(-\epsilon)$ reflects the emission spectrum of photons with energy $|\epsilon|$ showing a sharp kink at $eV$ and a substantial over-bias emission. The parameters are the resonance frequency $\omega_{0}=1.1eV$, the broadening $\eta=0.2eV$, and the coupling parameter $g_{c}z_0^2=0.3$. See the text for further explanations of the parameters.}
\label{fig:overview}
\end{figure}

As a first step, we develop a model of the SPP-mediated light emission of a tunnel junction, inspired by the theory of environmental Coulomb blockade developed more than two decades ago in mesoscopic physics \cite{Ingold:92}. In this picture, tunneling in a junction is strongly modified because the electronic environment leads to fluctuations of the voltage difference across the junction, showing up as e.g.~a zero-bias anomaly in the differential conductance \cite{Devoret:90,Girvin:90}. Considering that non-symmetrized current fluctuations induced by the combined tunnel junction-environmental impedance system can be seen as light emission \cite{Lesovik:97,Gavish:00,Aguado:00}, we suggest to model the SPP resonance as an electromagnetic resonator with appropriate damping interacting with the tunneling electrons of the STM (see Fig.~\ref{fig:overview}). 

In the second step, to theoretically address the problem we will make use of a recent observation in Ref.~\cite{Tobiska:06} stating that phase fluctuations in a coherent conductor-environment system lead to a subtle interference effect between two-photon processes and two-electron processes that may be identified by a strongly coupled quantum tunneling detector.
We will adapt the formalism of Ref. \cite{Tobiska:06} to the quantum detection of light emission from our coupled junction-resonator system,
but since experimentally the detector is far away from the junction and the emission efficiency is only $\lesssim 10^{-4}$ \cite{Schull:09}, it is sufficient to work in lowest order in the detector coupling, $\alpha$, in our model, see Fig.~\ref{fig:overview}c.
However, as we will show below, it is absolutely essential that the tunnel junction is described as a non-Gaussian quantum noise emitter, which at the same time means that the observed over-bias emission is a new manifestation of the nontrivial statistics of quantum transport \cite{QT:09,Reulet:03}. 
Using the method of \cite{Tobiska:06} we calculate the emission spectrum for all energies up to second order in the tunnel conductance and find quantitative agreement with experimental results of \cite{Schull:09}.

We will start by showing how we intend to model the interaction between the tunneling current and the SPP using methods of environmental Coulomb blockade theory \cite{Ingold:92,Devoret:90}.
According to standard theory \cite{Ingold:92,Devoret:90}, we model the tunneling from the STM tip to the surface in an electromagnetic environment as the circuit diagram depicted in Fig.~\ref{fig:overview}c. We consider a tunnel conductor with a dimensionless conductance $g_{c}=R_{Q}/R_{c}$ with $R_Q=h/2e^2$ and $R_c$ being the quantum and tunneling resistances, respectively. The junction is coupled to a damped LC circuit, which we model by an impedance $z_\omega=iz_{0}\omega \omega_{0}/(\omega_{0}^2-\omega^2+i\omega \eta)$, where $\omega_0=1/\sqrt{LC}$ is the resonance frequency of the SPP mode, $\eta=1/RC$ models the damping and $z_0=\sqrt{L/C}/R_{Q}$. We will later determine these parameters from the experiment \cite{Schull:09}. The interaction between the tunnel junction and the SPP occurs in this model via the dynamical voltage fluctuations on the node between the tunnel junction and the LRC circuit, which can be expressed by the phase variable $\varphi(t)=\frac{e}{\hbar}\int ^t_{-\infty} dt V(t')$. 

To model the emission detection of the photons, we follow the standard path and model the detector as a two-level system, in which the emitted photons trigger transitions between states characterized by an energy difference $\epsilon$ and a matrix element $\mathcal{T}$. We introduce a coupling constant $\alpha$ between the voltage fluctuations and the energy level of the detector, viz. $\epsilon\to\epsilon+\alpha eV(t)$. Finally, we will take the interaction to be weak, since the photon detectors in the real experiments are far away from the junction. Using Fermi's golden rule and setting $\hbar=1$, the detection rate at energy $\epsilon$  due to the fluctuations  of $\alpha \varphi(t)$ \cite{Ingold:92,Devoret:90,Tobiska:06} is
\begin{equation}
	\label{eq:rate}
	\Gamma(\epsilon)=\lvert \mathcal{T} \rvert ^2 \int dt 
	\langle e^{i\alpha \varphi(t)} e^{-i\alpha \varphi(0)} \rangle e^{i\epsilon t}.
\end{equation} 
To calculate $\langle e^{i\alpha \varphi(t)}e^{-i\alpha
\varphi(0)}\rangle$, we employ the path integral method, in which the real fields $\varphi^\pm(t)$ are defined on the forward and backward Keldysh contours, respectively. The dynamics of the coupled SPP-electron system is determined 
by the Keldysh actions of the conductor, $\mathbb{S}_{c}$, and the circuit, $\mathbb{S}_{e}$.
The correlator can then be represented as
\begin{eqnarray}
	\langle e^{i\alpha \varphi(t)}e^{-i\alpha \varphi(0)}\rangle=& \int
	\mathcal {D}[\Phi]\exp \{
	-i\mathbb{S}_{e}[\Phi]-i\mathbb{S}_{c}[\Phi]  \nonumber \\
	&+i\alpha[-\varphi^{+}(0)+\varphi^{-}(t)]\},
\end{eqnarray}
where $\Phi=((\varphi^++\varphi^-)/2, \varphi^+-\varphi^-)^T$. The action of the LRC circuit, i.e. the damped LC oscillator acting as the environment on the tunnel conductor, is quadratic in the fields and at zero temperature given by \cite{Kindermann:03,Kamenev:01}
\begin{equation}
\mathbb{S}_{e}=\int d\omega \Phi^{T}_{-\omega}A_\omega\Phi_{\omega}\,
,\,
A_\omega=-\frac{i}{2}\left( \begin{array}{cc} 0 &
-\frac{\omega}{z_{-\omega}} \\
\frac{\omega}{z_{\omega}} & \vert \omega \vert \Re \{
\frac{1}{z_{\omega}} \} \end{array} \right). \nonumber
\end{equation}
The action $\mathbb{S}_{c}$ can be expressed in terms of Keldysh Green's functions $\check{G}_{L,R}$ for the free electrons on the left ($L$) and right ($R$) sides of the tunneling barrier:  
\begin{equation}
\mathbb{S}_{c}=\frac{i}{8} g_{c}\int dtdt'\mathrm{Tr}\{ \check{
G}_{L}(t,t'), \check{G}_{R}(t'-t) \}\,.
\end{equation}
With  the help of the equilibrium Keldysh Green's function
\begin{equation}
	\check{G}(\omega)=\left(
	\begin{array}{cc}
	1-2f(\omega) & 2f(\omega) \\
	2[1-f(\omega)] & 2f(\omega)-1
	\end{array}\right) , \nonumber  
\end{equation} 
containing the Fermi function $f(\omega)=[\exp(\beta \omega)+1]^{-1}$, we can write $\check G_R(\omega)=G(\omega-eV)$ and hence $\check G(t)=\int d\omega \exp(-i\omega t)\check G(\omega)/2\pi$.
Again using the Fourier representation, we write $\check{G}_{L}(t,t^{\prime }) =\check U^\dagger(t) \check{G}(t-t^{\prime })\check U(t')$ with the counting fields introduced as \cite{Belzig:01}
\begin{equation}
\check U(t)=\left(
\begin{array}{cc}
e^{-i\varphi ^{+}(t)} & 0 \\
0 & e^{-i\varphi ^{-}(t)}
\end{array}
\right). \nonumber
\end{equation}
This concludes the description of our theoretical formalism. Unfortunately, the rate cannot be calculated exactly since the action of the conductor is non-Gaussian and we need an approximation scheme.

A simple approximation is considering only the Gaussian part of the conductor action.
In this case, the whole path integral becomes Gaussian and corresponds to the well-known results from P(E) theory. The quadratic part of the conductor action reads
\begin{equation*}
	\mathbb{S}_{c}^{\mathrm{G}} 
	= \int d\omega \Phi_{-\omega}^{T} B_{\omega} \Phi_{\omega}\,,\,
	B_\omega= -\frac{i}{2} \left( \begin{array}{cc}
	0 & -\omega g_{c} \\
	\omega g_{c} & S_{c}(\omega)
	\end{array} \right),
\end{equation*} 
with the symmetrized quantum noise of a tunnel contact $S_{c}(\omega)=g_{c}(\lvert \omega \rvert+Y(|\omega|-eV))$ using $Y(\omega)=-\omega \theta (-\omega)$. We will later discuss quantitative limitations of this approximation. However, already now we see that the Gaussian part alone will only lead to single photon emission with sub-bias energies.

\begin{figure}[t]
\centering
\includegraphics[width=0.9\columnwidth,clip=true]{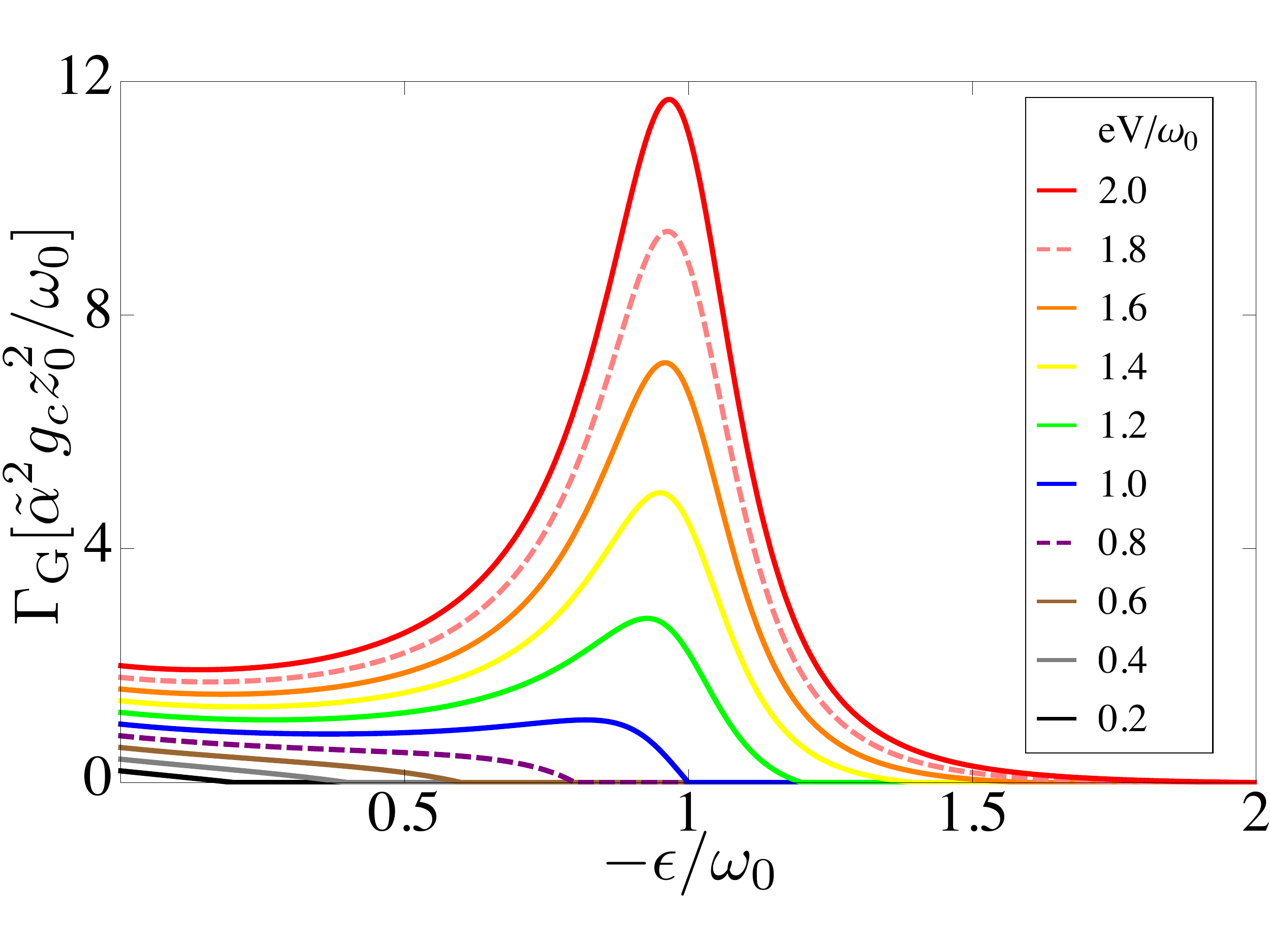}
\caption{The Gaussian contribution to the emission spectrum for different bias voltages. The SPP peak becomes clearly visible as the bias voltage exceeds the resonance energy, $\omega_0$. 
In all cases, the spectrum sharply drops to zero for $-\epsilon>eV$. This behavior ascertains that the responsible processes are limited by single-electron tunneling events. The broadening parameter is chosen as $\eta=0.3\omega_{0}$.}
\label{fig:2}
\end{figure}

Combining all the quadratic parts from both the LRC circuit and the conductor in a single matrix $D_{\omega} \equiv A_{\omega}+B_{\omega}$, the correlation function $\langle e^{i\alpha \varphi(t)}e^{-i\alpha \varphi(0)} \rangle\equiv e^{\alpha^2J(t)} $ can be evaluated. As a result, one finds
\begin{eqnarray}
	J(t) = 
	\int d\omega \frac{\lvert \tilde{z}_{\omega} \rvert ^{2}}{\omega^{2}}
	 S_t(\omega)(e^{-i\omega t}-1),
\end{eqnarray}
where $S_t(\omega)=S_{c}(\omega)+g_{c}\omega+2Y(-\omega)\Re \{ 1/z_{\omega}\}$  is the total noise spectral density and the impedance $\tilde{z}_{\omega}=z_{\omega}/(1+z_{\omega}g_{c})$. The renormalized impedance $\tilde z_\omega$ is the parallel connection of the tunnel junction and the environmental impedance as seen by the detector. This means the factor $g_c$ leads to an increased damping and can be absorbed in a renormalized $\eta$. 
From Eq. (\ref{eq:rate}), in lowest order in $\alpha^2$, we find the rate
\begin{equation}
	\label{eq:rategauss}
	\Gamma_{\mathrm{G}}(\epsilon) =\tilde\alpha^2
		\frac{\lvert \tilde{z}_{\epsilon} \rvert^{2}}{\epsilon^{2}}
		\left[   g_c Y(|\epsilon|-eV) +  \Re \left(\frac{2}{\tilde z_{\epsilon}}\right)Y(-\epsilon) \right] .
\end{equation}
Here, we have introduced a prefactor $\tilde\alpha^2=\lvert \mathcal{T} \rvert^2 \alpha^2 $.

\begin{figure}[t]
	\centering
	\includegraphics[width=0.9\columnwidth,clip=true]{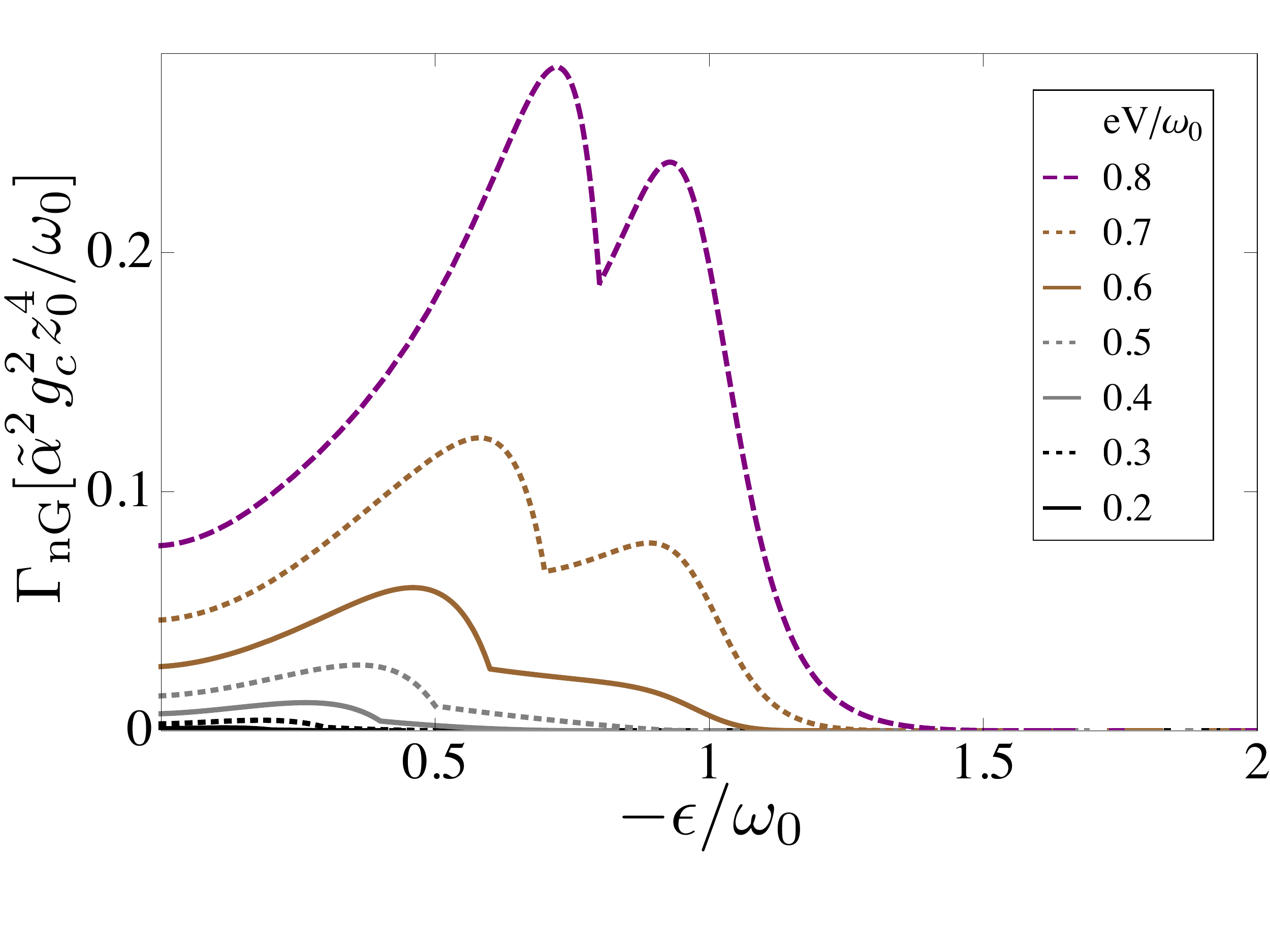}
	\caption{The non-Gaussian emission spectrum for  different bias voltages. The spectrum is clearly induced by the SPP resonance and shows a kink at the bias voltage. The over-bias emission rate is distinctly visible and the scaling with $g_c^2$ shows that this effect is due to two-electron tunneling processes. The broadening of the resonance is $\eta=0.3\omega_{0}$.}
	\label{fig:3}	
\end{figure}

The result (\ref{eq:rategauss}) matches the simple expectation from the golden rule \cite{Gavish:00}. The photon emission, that is described by energies $\epsilon<0$, is only caused by the non-equilibrium electrons of the tunnel junction and is therefore limited by the maximum energy $eV$ available for inelastic transitions. As the emission requires interaction with the environmental resonator, the electronic inelastic emission spectrum is filtered by the SPP resonance. This is demonstrated in Fig.~\ref{fig:2}, which shows the emission rate for different bias voltages. There is always a sharp threshold for $-\epsilon=eV$ and the SPP resonance becomes visible if the threshold is larger than the resonance energy, $\omega_0$. 

We have seen that the Gaussian approximation of the rate does not result in an over-bias emission. Hence, we have to take the non-Gaussian statistics of the tunnel junction into account. It is not possible to calculate that part exactly. Fortunately, we can make use of the limit $z_\omega^2 g_c\ll1$ motivated by the experiment by Schull and co-workers \cite{Schull:09}. From our fitting later, we can infer that the fluctuations of the phase are small due to the dominating Gaussian part governed by the small environmental impedance.

Therefore, we take the non-quadratic part from the higher order expansion of $\mathbb{S}_{c}$ in $\Phi$  into account as $\mathbb{S}_c=\mathbb{S}_c^G+\mathbb{S}_c^{(3)}+\mathbb{S}_c^{(4)}+O(\Phi^5)$. The Gaussian part of the action can be combined with the environmental action, i.e $\mathbb{S}_c^G+\mathbb{S}_e\rightarrow \mathbb{S}_c^G$. Due to the above assumptions, the remaining terms are small and we can make the expansion $\exp[-i\mathbb{S}_c^{(3)}-i\mathbb{S}_c^{(4)}]\approx1-i\mathbb{S}_c^{(3)}-i\mathbb{S}_c^{(4)}$. This approximation is possible since the Gaussian part of the action is dominated by the fluctuations of the small impedance of the environment, viz. $\Phi^2< z_\omega/\omega^2$, and therefore the higher order terms are small by the factor $g_cz_\omega^2 \ll1$. After the expansion, the remaining path integral is just the Gaussian average of the third and fourth moments. The Gaussian average is then given by
\begin{equation}
	\langle\langle \cdots \rangle\rangle 
	\equiv \int \mathcal{D} [\Phi] (\cdots) 
	e^{\int d \omega \{-i \Phi ^{T}_{-\omega} D_\omega \Phi_{\omega}
	+i \alpha b^{T}_{\omega}(t)	\Phi_{\omega}\}}\,,
\end{equation}
where $b_{\omega}(t)=(e^{-i\omega t}-1, -(e^{-i\omega t}+1)/2)^{T}$.
Now, all remaining averages can be calculated using Wick's theorem and, as usual, this gives the sum over all possible pairings of single and double averages.
The basic averages in frequency space can preferably be expressed
in terms of the building blocks  $D_\omega$ and $b_{\omega}(t)$:
\begin{align}
	\langle\langle \Phi_{\omega} \rangle\rangle & 
	= \frac{\alpha}{2}D^{-1}_{\omega}b_{-\omega}^{\phantom{-1}}(t) e^{\alpha^2 J(t)}
	\\
	\langle\langle \Phi_{\omega} \Phi_{-\omega}^{T} \rangle\rangle &
	= -\frac{i}{2}D_{\omega}^{-1} e^{\alpha^2 J(t)}\,.
\end{align}
Note that these expressions still are valid for an arbitrary value of $\alpha$.
A drastic simplification arises if we limit ourselves to the experimentally relevant weak detection limit in which $\alpha\ll1$. The leading order contributions to the detector rate are given by combinations of the type $\langle\langle \varphi_{\omega} \rangle\rangle \langle\langle \varphi_{-\omega} \rangle\rangle \langle\langle \varphi_{\omega'}\varphi_{-\omega'} \rangle\rangle$ since single averages are of leading order $\alpha$. Contributions of zeroth order in $\alpha$ are time independent and therefore only play a part in the elastic rate characterized by $\epsilon=0$, which is not of interest here. Limiting ourselves to the light emission, i.e. $\epsilon < 0$, we find 
\begin{widetext}
\begin{eqnarray}
	\Gamma_{\mathrm{nG}} (\epsilon<0)&=&\frac{\tilde{\alpha}^2}{8} g_{c}^2 \frac{\lvert \tilde{z}_{\epsilon} \rvert^{2}}{\epsilon^{2}} \left\{ \int^{eV}_{0} d \omega  \frac{\lvert \tilde{z}_{\omega} \rvert^{2}}{\omega^{2}} (eV-\omega) [ \xi(\omega+\epsilon)+\xi(\omega-\epsilon)+2\epsilon-\xi(\epsilon) ]\right. \nonumber  \\ \label{eq:non-Gaussian}
&+&\frac{2 Y(-eV-\epsilon)}{\epsilon} \int^{\infty}_{0} \frac{d \omega}{\omega} \Big( \Re\{ \tilde{z}_{\epsilon} \} \Re\{ \tilde{z}_{\omega}\} [ \xi(\omega+\epsilon)-\xi(\omega-\epsilon)-2\epsilon-\xi(\epsilon) ] \\ \nonumber
&+&\Im\{ \tilde{z}_{\epsilon}\}\Im\{ \tilde{z}_{\omega}\} [ 4eV+\xi(\omega+\epsilon)+\xi(\omega-\epsilon) -2 \xi(\omega)-2\xi(\epsilon) ] \Big) \Bigg\}
\end{eqnarray}
\end{widetext}
with $\xi(\omega)=\lvert \omega+eV \rvert+\lvert \omega-eV \rvert$. This is the main result of our work and describes the influence of the non-Gaussian contribution to the light emission in the whole energy range. Note that it can be further simplified in the over-bias regime for $eV<-\epsilon<2eV$ and takes the same form as in Ref.~\cite{Tobiska:06} to order $\alpha^2$. We also see that the over-bias emission rate is $\sim g_c^2$, which signals the fact that a correlated two-electron tunneling process is responsible.

The non-Gaussian rate (\ref{eq:non-Gaussian}) explains the emission of photons with energies $-\epsilon>eV$.
The detailed behavior of this rate as a function of energy is shown in  Fig.~\ref{fig:3} for different values of $eV$.
We observe that the rate has a distinct kink for $-\epsilon=eV$, which can be seen as a signature of the sharp Fermi edge. This leads, for $eV<\omega_0$,
to a two-peak structure with peaks of comparable heights above and below the threshold voltage. For higher voltages, only a single peak at the resonance frequency remains.  

\begin{figure}[tb]
	\centering
	\includegraphics[scale=0.35]{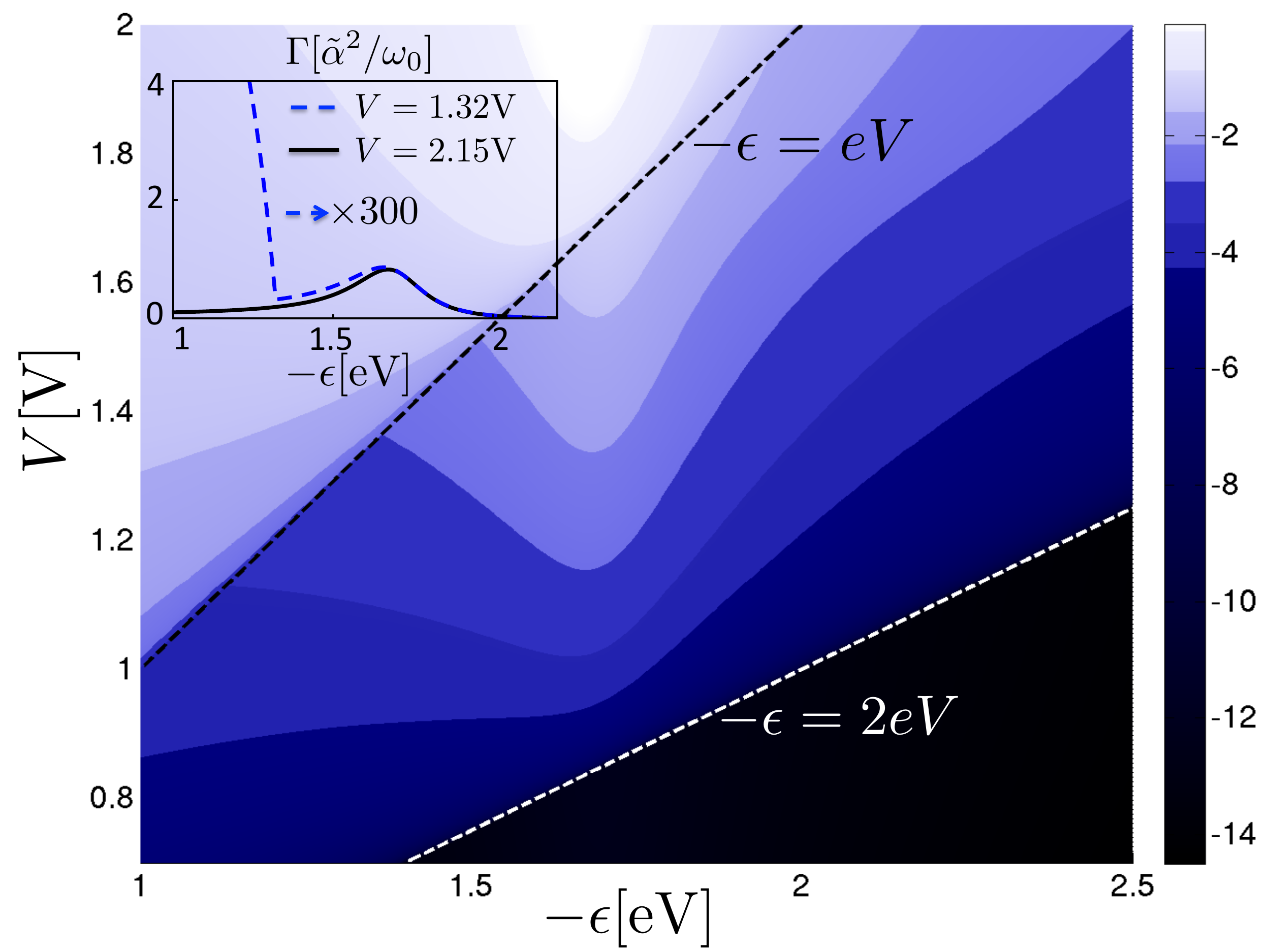}
	\caption{Main: Light emission spectrum on a logarithmic scale as a function of bias voltage. The SPP resonance energy is taken to be on the order of the experimental value $\omega_{0}=1.7$ eV, $g_cz_0^2=0.1$ and the broadening  is taken to be $\eta=0.2\omega_0$.
	The one- and two-electron thresholds at $-\epsilon=eV$ and $=2eV$ are indicated by dashed lines. Inset: To extract the coupling parameter $g_cz_0^2=0.1$, we compare the peak values at $-\epsilon=\omega_0$ for two different bias voltages: $V$=2.15 V (solid line); $V$=1.32 V (dashed line). By scaling the low-bias curve by a factor 300, we find curves similar to those of Fig.~2a in Ref.~\cite{Schull:09}. Note that we have taken $\eta=0.3\omega_0$ to achieve a better agreement of the resonance shape with the experiment.} 
	\label{fig:fullrate}
\end{figure}

To compare our theoretical model with the experimental data \cite{Schull:09}, we have to take the Gaussian as well as the non-Gaussian rates into account. As mentioned above, the two rates differ parametrically by a factor of $g_cz_0^2$. In fact, we can determine this parameter by comparison with the experimental results. In the inset of Fig.~\ref{fig:fullrate}, we show the total rate $\Gamma=\Gamma_{\mathrm{G}}+\Gamma_{\mathrm{nG}}$ for two different bias voltages. These rates have to be compared to the results prestented in Fig.~2a of Ref.~\cite{Schull:09}. From the relative scaling of the two curves by a factor of 300 and the width of the resonance, we determine the parameters $g_cz_0^2\approx 0.1$ and $\eta\approx 0.3\omega_0$, respectively. Note that the experimental results depend on the detailed surroundings of the STM tip's position. Using these parameters, we show the full bias-voltage and energy-dependent emission rate on a logarithmic scale in the main panel of Fig.~\ref{fig:fullrate}. The comparison to Fig.~1b of Ref.~\cite{Schull:09} is striking although the resonance parameters in the experiment are different. We clearly observe the threshold behaviors at $-\epsilon=eV$ and $-\epsilon=2eV$. Recently the light in the $2eV$ energy range has been investigated in more detail experimentally \cite{Schneider:13}, but a confirmation of a well-developed threshold behavior still needs more evidence.  We should add that experimentally the data are cut for $-\epsilon<1.2eV$, which is attributed to the detector sensitivity. Finally, we should emphasize that the experimental finding that the one-(two-)electron rate scales approximately with $g_c(g_c^2)$ is correctly reproduced by our theoretical model.

In conclusion, motivated by the experimental observation of photons with over-bias energies emitted by tunnel junctions,
we have developed a model of electron-SPP interaction based on dynamical Coulomb blockade. The interplay between the non-Gaussian statistics of the tunneling process and the resonant excitations of the SPP leads to a pronounced emission spectrum in which the SPP spectrum is overlaid with the sharp quantum threshold behavior determined by the bias voltage $eV$. Furthermore, the theory reproduces the experimentally observed emission with energies larger than the single-particle energy $eV$. A comparison of our model calculation to the experimental spectrum reveals a quantitative agreement of both the spectrum of the SPP resonance and the quantum thresholds. Our research enables a new level of modeling electron-SPP interaction in nano-size contacts. Furthermore our calculation shows that the over-bias emission can be used to experimentally probe higher-order quantum fluctuations from a tunnel junction.
Open questions concern going beyond the tunneling approximation and the weak coupling regime or considering the effect of molecules in the junction \cite{Dong:04}.

{\it Acknowledgments}.
We acknowledge useful discussions with R. Berndt, F. Haupt, K. Kaasbjerg, A. Nitzan and N. Schneider. This work was supported by the DFG through SFB 767 and by the Kurt-Lion-Foundation.

\end{document}